\documentclass[aps,pre,showpacs,twocolumn,amssymb,floats,epsfig]{revtex4}

\usepackage{graphicx}
\usepackage{epsfig}
\usepackage{amsfonts}
\usepackage{amsmath}
\usepackage{amssymb}
\usepackage{subfigure}
\usepackage{color}

\newcommand{\ct}{\cite}
\newcommand{\bi}{\bibitem}
\newcommand{\be}{\begin{equation}}
\newcommand{\ee}{\end{equation}}
\newcommand{\ba}{\begin{eqnarray}}
\newcommand{\ea}{\end{eqnarray}}

\newcommand{\non}{\nonumber}

\newcommand{\de}{\delta}
\newcommand{\la}{\lambda}

\begin{document}
\title{Survival probability in a quenched Majorana chain with an impurity}
\author{Atanu Rajak}
\affiliation{CMP Division, Saha Institute of Nuclear Physics, 1/AF Bidhannagar, Kolkata 700 064, India}
\affiliation{Department of Physics, Jack and Pearl Resnick Institute, Bar-Ilan University, Ramat-Gan 52900, Israel}
\author{Tanay Nag}
\affiliation{Department of Physics, Indian Institute of Technology Kanpur, Kanpur 208 016, India}

\begin{abstract}

We investigate the dynamics of  a one-dimensional $p$-wave superconductor with next-nearest-neighbor hopping and 
superconducting interaction derived from  a three-spin interacting Ising model in transverse field by mapping 
to Majorana fermions. The next-nearest-neighbor hopping term  leads  a new topological
phase containing two zero-energy Majorana modes at each end of an open chain, compared to a nearest-neighbor $p$-wave 
superconducting chain. 
We study the Majorana survival probability (MSP) of a particular Majorana edge state when 
the initial Hamiltonian ($H_i$) is changed to the quantum critical as well as off-critical final  Hamiltonian ($H_f$) 
which additionally contains an impurity term ($H_{imp}$) that breaks the time-reversal invariance. 
For the off-critical quenching  inside the new topological phase with $H_f= H_i +H_{imp}$, and small 
impurity strength ($\la_d$), we observe a perfect oscillation of the MSP as a function of time with a single frequency 
(determined by the impurity  strength $\la_d$) that can be analyzed from an equivalent two-level problem.
On the other hand, the MSP shows a  beating like structure with time for quenching to the phase boundary 
separating the topological phase (with two edge Majoranas at each edge) and the non-topological phase 
where the  additional frequency is given by inverse of the system size. We attribute this behavior of the MSP to 
the modification of the energy levels of the final Hamiltonian due to the application of the impurity term.


\end{abstract}
\maketitle

\section{Introduction}
In the rapidly growing field of research in topological quantum computation, quantum information processing 
and decoherence, Majorana fermions, introduced by Ettore Majorana~\ct{majorana37} in the context of the 
existence of real solution of Dirac equation,
play a key role \ct{kitaev09,budich12,schmidt12,tewari07}. 
The theoretical prediction of the existence of zero-energy Majorana edge modes in an open chain
has paved the way for understanding of various topological character for a system like one-dimensional 
spinless $p$-wave superconductor~\ct{kitaev01,fulga11,sau12,lutchyn11,degottardi11,degottardi13,thakurathi13,wdegottardi13,alicea12}.  
An interesting proposal has been made for achieving the Majorana states by the proximity effect between 
the surface state of a strong topological insulator and a $s$-wave superconductor~\ct{fu08}. 
A topological phase is characterized by a topological invariant number, for example, the number of 
zero-energy Majorana edge modes in the case of $p$-wave superconducting chain. This number does 
not change unless the system goes from one topological phase to the other separated by a 
quantum critical line, as also happens in a topological insulators~\ct{hasan10,qi11}.
The zero-energy Majorana modes have recently been found experimentally in nanowires coupled to 
superconductors~\ct{Kouwenhoven12,deng12,das12,chang13,lee14}; although there exists some theoretical 
contradictions with the experimental observation ~\ct{rainis13}.
An experimental realization of the hybridization of Majorana fermions has also been observed through the 
zero-bias anomalies in the differential conductance of an InAs nanowire coupled to superconductor \ct{Finck13}.


On the other hand, given the recent interest in the non-equilibrium quenching dynamics of quantum many body
systems across QCPs, the investigations of different topological systems in out-of-equilibrium
have become a prime research field~\ct{dutta15,bermudez09,bermudez10,wang14,setiawan15,wu15,sacramento14,
sacramento15,dai15,zvyagin15,hegde15,wang15,bhattacharya16,lee16}. In this connection, the dynamics of an edge state  has been 
extensively investigated in one \ct{rajak14} and two dimensional~\ct{patel13} topological systems.
Recently, the dynamics of the Majorana edge state also has been studied 
following a sequence of quenches in an one-dimensional system ~\ct{sacramento16}.
It is noteworthy to mention that the dynamical generation~\ct{thakurathi13}, 
formation and manipulation \ct{perfetto13} of Majorana edge states for a driven system have also 
studied extensively. Additionally, the theoretical prediction of an adiabatic transport of an edge Majorana through  
an extended gapless region has been made in a $p$-wave superconducting chain with complex hopping 
term~\ct{rajak14b}. 

We consider here a generalized one-dimensional Ising model in a transverse field with a 
three-spin interaction term~\ct{kopp05} which can be written in term of fermionic operators using 
the Jordan-Wigner transformation. This longer-range interacting model has a 
richer phase diagram containing an extra topological phase with two zero-energy 
Majorana edge modes~\ct{niu12} as compared to the one-dimensional $p$-wave superconductor 
which is fermionized version of the transverse field XY model~\ct{thakurathi13}. 
If we break the time-reversal symmetry of the above Hamiltonian by adding an
impurity term, the Majorana modes at one end of the chain vanishes in the topological 
phase with two Majorana modes depending on the nature of the impurity term~\ct{niu12}. 
Our main aim here is to investigate the fate of a Majorana edge state 
under time evolution governed by the Hamiltonian with an impurity that destroys the edge 
state in equilibrium.


In particular, we investigate the Majorana survival probability (MSP) as a function of time 
following both critical and off-critical quenches  in the presence of an impurity term in the final Hamiltonian. 
We show that this impurity term modifies the energy levels of the final Hamiltonian and consequently 
an extra time scale appears in the system which is reflected in the behavior of 
the MSP. Most Interestingly, for weak impurity strength, we find that MSP exhibits  perfect temporal oscillation 
with a single frequency following an off-critical quench inside the topological phase carrying two edge Majoranas; the 
oscillation frequency is determined by the inverse of impurity strength.
In this connection, we note that recently the single frequency oscillation in entanglement spectrum 
also has been found for a two-impurity Kondo model by suddenly changing the RKKY interaction strength~\ct{bayat15}.
On the other hand, MSP shows a beating like structure when the system 
is quenched up to a QCP (i.e., the phase boundary); the other energy scale at the QCP is the inverse of the system size.
To the best of our knowledge, it is the first work that suggests that there could be a nearly perfect
oscillation in the MSP even for the off-critical quenching. The above mentioned behaviors
of the MSP are destroyed when impurity strength is appreciably large.

This paper is organized as follows: In Sec.~\ref{model} we introduce the three-spin interacting 
transverse field Ising model and discuss its phase diagram. We also mention 
the effect of an impurity term on the different phases of the model.
In Sec.~\ref{qd}, we define the quantity MSP which has been calculated 
for different sudden quenches.
In the subsequent subsections, we illustrate our results 
of the  MSP under an application of the impurity term for critical as well as 
off-critical quenches. 
Finally, we provide our concluding remarks in Sec.~\ref{conclusion}.

\section{Model}
\label{model}
\begin{figure}[ht]
\begin{center}
\includegraphics[height=2.4in]{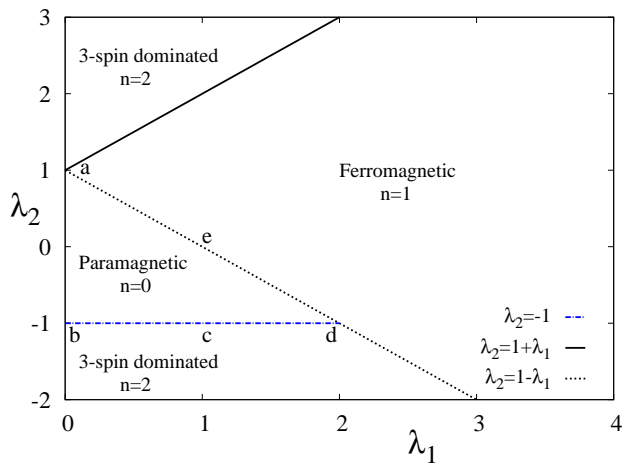}
\end{center}
\caption{(Color online) Phase diagram of the model with Hamiltonian in Eq.~(\ref{ham1}) for $h=1$. 
The phase boundary $\la_2=1+\la_1$ separates upper topological phase (with two zero energy Majorana 
edge states at each end) from one Majorana ($n=1$) topological phase. The other two phase 
boundaries are $\la_2=1-\la_1$ 
($a-e-d$ line) and $\la_2=-1$ ($b-d$ line). The paramagnetic region corresponds to a non-topological phase with
 $n=0$ Majorana modes. Both the topological 
phases with $n=2$ Majorana modes i.e., lower and upper, are characterized by the presence of $a_1$ and  
$a_2$ isolated modes at the left end and $b_{N-1}$, 
$b_N$ at the right end of the chain.}
\label{pd}
\end{figure}

The Hamiltonian of a three-spin interacting transverse Ising model with $N$ spins is given by~~\ct{kopp05}
\be
H~=~-\sum_n(h\sigma_n^z~+~\la_1\sigma_n^x\sigma_{n+1}^x~+~\la_2\sigma_{n-1}^x\sigma_{n}^z\sigma_{n+1}^x),
\label{ham1}
\ee
where $h$, $\la_1$ and $\la_2$ are transverse magnetic field, cooperative interaction and 
three-spin interaction respectively. $\sigma^{\alpha}$ ($\alpha=x,y,z$) are the standard 
Pauli matrices. This model can be exactly solved by Jordan-Wigner (JW) transformation~\ct{lieb61}
by mapping the spins into the spinless fermions. 
The JW transformation is defined as
\begin{eqnarray}
\sigma_n^-&=&\prod_{j=1}^{n-1}\left(-\sigma_j^z\right)c_n,\nonumber\\
\sigma_n^z&=& 2c_n^{\dagger}c_n-1,
\label{jwt}
\end{eqnarray}
where $\sigma_n^{\pm}=(\sigma_n^x \pm i\sigma_n^y)/2$, and $c_n^{\dagger}$, $c_n$ are the fermionic
creation and annihilation operators respectively. In terms of the JW fermions,
the Hamiltonian in Eq.~(\ref{ham1}) is given by
\ba
H&=&-\sum_{n=1}^{N}\Big[h(2c_n^\dagger
c_n-1)-\lambda_1(c_n^\dagger c_{n+1}+c_n^\dagger
c_{n+1}^\dagger+H.c.)\non \\
&-&\lambda_2(c_{n+1}^\dagger
c_{n-1}^{\dagger}-c_{n-1}^{\dagger}c_{n+1}+H.c.)\Big].
\label{f_ham}
\ea
The three-spin interacting term in the Hamiltonian~(\ref{ham1}) gives rise to the next-nearest-neighbor 
hopping and superconducting gap terms  in addition to the nearest-neighbor 
hopping and the superconducting gap terms. 
The Hamiltonian (\ref{f_ham}) reduces to a direct sum of $2 \times 2$ decoupled Hamiltonian $H_k$ in 
momentum space under the periodic boundary condition. In the momentum space representation 
the Hamiltonian (\ref{f_ham}) is given by 
\ba
H&=&\sum_{k>0}~(c_k^{\dagger}~ c_{-k})\hspace{2mm}H_k\hspace{2mm}{c_k\choose~c_{-k}^{\dagger}}, ~ {\rm with} \non \\
H_k&=& (h+\la_1 \cos k -\la_2\cos 2k)\sigma^z+(\la_1\sin k \non\\
&-&\la_2 \sin 2k)\sigma^x
\label{hma_k}
\ea 
where $c_k=(1/\sqrt{N})\sum_n e^{-ikn} c_n$. 
The Hamiltonian in Eq.~(\ref{hma_k}) is diagonalized by a Bogoliubov transformation to 
obtain the energy spectrum of the system, given by 
\be
\varepsilon_{k}=\pm 2\sqrt{h^2+\lambda_{1}^{2}+\lambda_{2}^{2}+2\lambda_{1}(h-\lambda_{2})\cos k -2h\lambda_{2}\cos 2k}.
\label{gap}
\ee
The Hamiltonian in Eq.~(\ref{ham1}) reduces to the transverse Ising model when $\la_2=0$; this model has a
quantum phase transition at $\la_1=h$ between a ferromagnetic and a paramagnetic phase where 
the energy gap vanishes for the critical modes $k_c= \pi$.
We set $h=1$ throughout the paper.
In order to investigate the phase diagram, as shown in Fig.~(\ref{pd}),  of the Hamiltonian (\ref{ham1}) one has to analyze the energy 
spectrum in Eq.~(\ref{gap}) as a function of $\la_1$ and $\la_2$.
It can be verified that the low energy excitation gap of the 
system vanishes on the quantum critical lines $\la_2=1+\la_1$ and $\la_2=1-\la_1$ for the 
critical modes $k=0$ and $\pi$, respectively. There is an another critical line
$\la_2=-1$, where the energy gap vanishes for $k=\cos^{-1}(-\la_1/2)$ implying that this transition can not occur for $\la_1>2$. The critical 
line $\la_2=1+\la_1$ corresponds to the phase boundary between the three-spin dominated 
and ferromagnetic phases. On the other hand, the critical line $\la_2=1-\la_1$ separates 
 the ferromagnetically ordered phase from the paramagnetic phase when $\la_2>-1$ and three-spin dominated phase for $\la_2<-1$.

In order to explore the topological properties of the model, we represent a Jordan-Wigner 
fermion $c_n$ in terms of the two Majorana fermions $a_n$ and $b_n$, where
\be
a_n = c_n^\dagger+c_n,
b_n=-i( c_n^\dagger-c_n). 
\ee
These Majorana fermions are real and satisfy the following relations 
$\{ a_m, a_n \} =
\{ b_m, b_n \} = 2 \de_{mn}$ and $\{ a_m, b_n \} = 0$. 
Using the Majorana operators, the Hamiltonian in Eq.~(\ref{f_ham}) with open boundary condition (OBC) 
can be re-written as
\be
H=-i\Big[-h\sum_{n=1}^Nb_na_n~+~\la_1\sum_{n=1}^{N-1}b_na_{n+1}~+~\la_2\sum_{n=2}^{N-1}b_{n-1}a_{n+1}\Big].
\label{ham2}
\ee
The three-spin interaction in Eq.~(\ref{ham1}) corresponds to a next-nearest-neighbor coupling in Majorana 
fermion operators that leads to an extra topological phase with two Majorana zero modes at each end of 
an open chain (see Fig.~\ref{pd}).
It can be easily shown using a special condition $h=\la_1=0$ in Hamiltonian (\ref{ham2}) that
the upper and lower three-spin dominated phases support two 
zero-energy Majorana edge modes at each end of the open chain: $a_1$ and $a_2$ exist at the left boundary, 
while, $b_N$ and $b_{N-1}$ are at the right boundary. On the other hand,
in the ferromagnetic region the open chain consists one Majorana edge mode at each end i.e., $a_1$ 
 at the left boundary and $b_N$ at the right boundary. 
The paramagnetic region does not have any topological 
property, hence no edge Majorana survives in the open chain. 

Let us now introduce some special terms in the Hamiltonian given in Eq.~(\ref{ham2})
that break the time-reversal symmetry (TRS) of the system. 
The total Hamiltonian containing such terms is given by
\be
H_T~=~-i\sum_m\sum_j[K_m a_ja_{j+m}+L_mb_jb_{j+m}]+H,
\label{trs_break}
\ee
where $K_m$ and $L_m$ are the real parameters. $m$ denotes the range of interaction. 
We note that these special terms can destroy multiple number of $a$ and $b$ Majorana edge modes depending on the 
non-zero values of $K_m$ and $L_m$, respectively. 
We here assume  a simplified situation $L_m=0$, $\forall$ $m$ and $K_m=\lambda_d$ for a single value of 
$m$, otherwise zero. Hence, the Eq.~(\ref{trs_break})
reduces to the form
\ba
&& H_T=H_{\rm imp}+H, {\rm where} \non \\
&&H_{\rm imp}~=~-i\la_da_ja_{q}, q=j+m.
\label{imp_ham}
\ea
Using JW transformation $H_{\rm imp}$ can be expressed in terms of spin operators
\be
H_{\rm imp}=\la_d\prod_{n=j+1}^{j+m-1}(-\sigma_n^z)\sigma_j^y\sigma_{j+m}^x.\non
\ee
For $m=1$, it reduces to $H_{\rm imp}=\la_d\sigma_j^y\sigma_{j+1}^x$ which represents the interaction between $y$ and $x$
components of two nearest-neighbor spins respectively. On the other hand, for larger $m$, it becomes non-local in terms of 
the spin operators. Nevertheless, in the Majorana language we can say it is quasi-local (see Eq.~(\ref{imp_ham})).

In the subsequent sections, we shall refer total Hamiltonian $H_T$ as $H$.
It has been found that the phases with two Majorana 
zero modes are affected by the impurity Hamiltonian, whereas 
the phase with one Majorana mode remains intact. Here, the impurity term is applied in the left end of the 
open chain, hence the Majorana modes at the 
left end (i.e., $a$ type) vanishes but the $b$ Majorana modes at the right end remain intact. 
On the other hand, if we apply the impurity in the bulk the coupling of $a_1$ 
and $a_2$ Majorana modes becomes very weak and they do not move away from zero energy.


\section{Quench dynamics of the chain and results}
\label{qd}

Here, we shall study the survival probability of an edge Majorana~\ct{rajak14} when the different 
terms of the Hamiltonian (\ref{ham2}) 
is quenched from one phase to the other or to the QCP separating two phases as shown in Fig.~1. 
To study the dynamics of the zero energy edge  Majorana mode after a sudden quench, we now define the Majorana 
survival probability (MSP) $P_m(t)$, defined as
\be
P_m(t)=\Big|\sum_{n=1}^{2N}|\langle\psi_m(\la_1,\la_2)|\Phi_n(\la'_1,\la'_2)\rangle|^2e^{-iE_nt}\Big|^2,
\label{prob}
\ee
where $|\psi_m(\la_1,\la_2)\rangle$ is an initial edge Majorana state for the parameters $\la_1$
 and $\la_2$, and $|\Phi_n(\la'_1,\la'_2)\rangle$ are 
the eigenstates of the final Hamiltonian with new parameters $\la'_1$, $\la'_2$ while  $E_n$'s 
are the eigenvalues of the final Hamiltonian. We study 
the dynamics of a zero energy edge state when the final Hamiltonian is either non-critical or
critical. As mentioned already we have kept the transverse field $h=1$ for all the quenching processes.

\begin{figure}[ht]
\begin{center}
\includegraphics[width=3.7in]{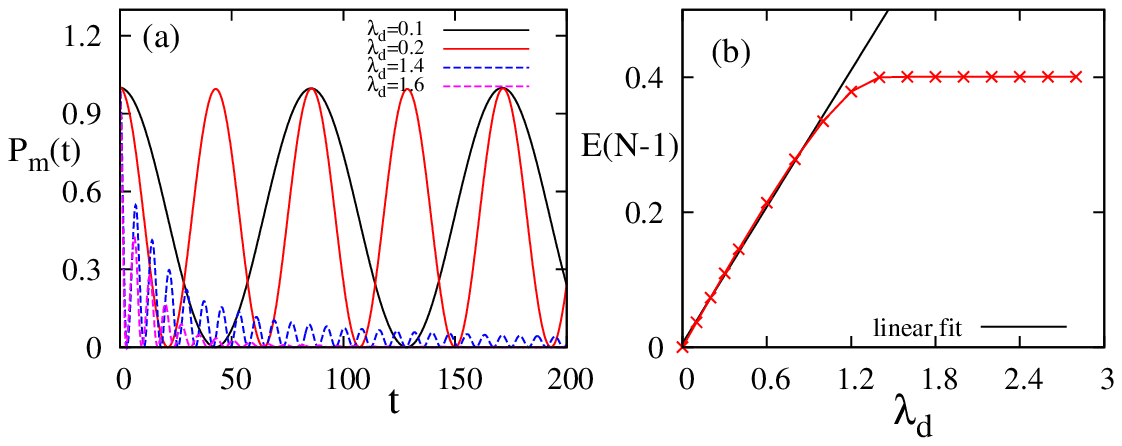}
\vskip -0.5cm
\includegraphics[width=3.7in]{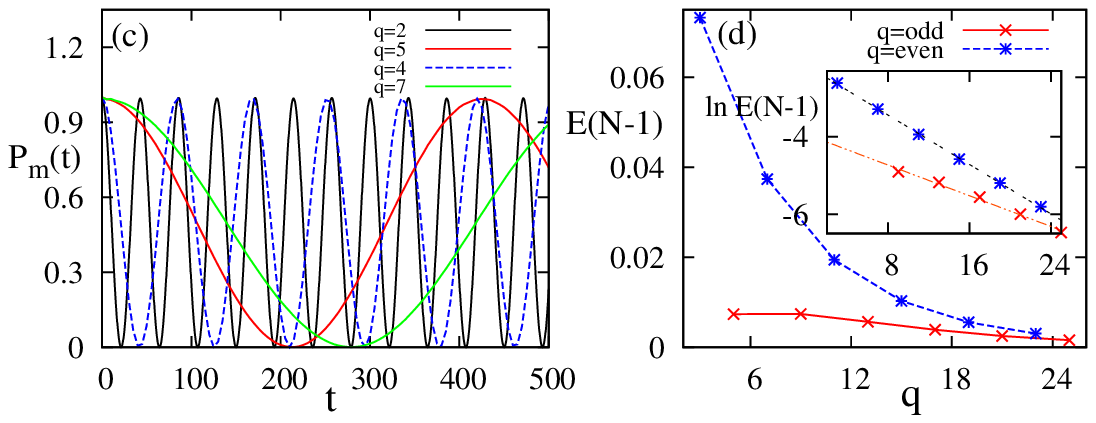}
\end{center}
\caption{(Color online) (a) The MSP of an initial zero energy edge Majorana ($a_1$)
shows perfect oscillation with time for smaller values of $\lambda_d$, whereas
it decays nearly equal to zero value and does not revive significantly for higher values of $\lambda_d$. 
The rate of decay increases as the value of $\la_d$ is increased. In this case, 
the initial Hamiltonian is considered to be in the $n=2$ phase with
$\la_1=0.2$ and $\la_2=2.0$, and the final Hamiltonian has an additional
 impurity term $H_{\rm imp}=\la_da_1 a_2$ over the initial Hamiltonian.
 Plot (b) indicates that the energy corresponding $a_1$ edge Majorana  
 increases linearly with $\la_d$ followed by a saturation as $\la_d$ becomes large.
Plot (c) shows the behavior of MSP for $a_1$ with $\la_d=0.2$ for different 
interaction range of impurity term $H_{\rm imp}=\la_da_1a_q$ where $q$ can take even or odd values.
Plot (d) indicates that the energy corresponding the $a_1$ Majorana decays exponentially
with $q$. On the other hand, inset shows that the prefactor 
of the exponential fall is different for even and odd $q$. Here, $N=200$.}
 \label{imp}
\end{figure}


\subsection{Non-critical quenching: MSP}
\label{result1}
We first discuss the effect of the impurity term (see Eq.~(\ref{imp_ham})) on the MSP (\ref{prob})
for the case of off-critical quenching. We here investigate the MSP after a quenching inside 
a phase or across a phase boundary while the impurity term is added only to the final Hamiltonian.

Let us begin with a situation when the system is quenched within the upper topological phase ($n=2$) 
by adding an impurity term.
For this quenching process the final Hamiltonian $H_f$ is simply given by initial Hamiltonian 
$H_i$ with the extra impurity term: $H_f=H_i+H_{imp}$.
 As discussed in Sec.~\ref{model}, 
the addition of such a term in Eq.~(\ref{ham2}) destroys the left end Majorana modes ($a_1$ and $a_2$) of an 
open chain. We are now interested to study the MSP in Eq.~(\ref{prob}) associated with $a_1$ Majorana mode following 
the above mentioned quench with $q=2$. 
As shown in Fig.~\ref{imp}(a), the MSP for $a_1$ edge Majorana shows perfect oscillations 
as a function of time $t$ for smaller values of $\la_d$, whereas, the damped oscillations are 
observed in the case of larger values of $\la_d$. A close observation suggests that for small $\la_d$, the time period of 
collapse and revival in the MSP is inversely proportional to $\la_d$.
Although, the Majorana zero mode $a_1$ is not present at the final Hamiltonian, it comes back 
periodically at the left end of the chain with time as an outcome of the sudden quenching.

We consider the overlap function $\alpha_n=|\langle\psi_m(\la_1,\la_2)|\Phi_n(\la'_1,\la'_2,\la_d)\rangle|$ of Eq.~(\ref{prob}) 
to analyze this off-critical oscillation of the MSP. 
We find that the overlap ($\alpha_n$) of an initial Majorana edge state with the $n$-th eigenstate 
of the final Hamiltonian for small $\la_d$ becomes non-zero for only two values of $n$, i.e., the states 
which deviate from zero energies due to the application of the impurity term. Hence
it is clear that the impurity term of smaller strength affects only two zero energy states 
corresponding to $a_1$ and $a_2$ Majorana modes and the other eigenstates remain unchanged 
which are indeed orthogonal to the initial edge Majorana resulting in zero overlaps. 

To analyze the off-critical oscillation quantitatively, we
would like to employ the degenerate perturbation theory. Let us consider the final
 Hamiltonian, $H_f=H_i + H_{\rm imp}$, where $H_{\rm imp}$ can be treated as the 
 perturbation. Then the first order correction to 
 the zero energy of two Majorana edge states
 becomes $E_{\pm}=1/2(W_{11}+W_{22}\pm \sqrt{(W_{11}-W_{22})^2+4|W_{12}|^2})$,
 with $W_{11}=\langle \psi_1| H_{\rm imp}|\psi_1\rangle$, $W_{22}=\langle \psi_2| H_{\rm imp}|\psi_2\rangle$ 
 and $W_{12}=\langle \psi_1| H_{\rm imp}|\psi_2\rangle$; $|\psi_{1/2}\rangle$ are two 
 wavefunctions of the $a$-type Majorana modes at the left end of the chain that
 are destroyed due to the application of $H_{\rm imp}$. In Majorana basis, the wavefunctions 
 of $a$-type Majorana edge modes are given by $\psi_n^T=(c_{n1},0,c_{n2},0,c_{n3},\cdots)$, where 
 $n=1,2$ and $c_{ni}\in \mathcal{R}$. Since these Majorana modes are localized at the end of the 
 chain, the value of $|c_{ni}|^2$ falls exponentially as $i$ is increased. Now, writing  $ H_{\rm imp}$ 
 as given in Eq.~(\ref{imp_ham}) in real space matrix form,
 it is easy  to show that $W_{11}=W_{22}=0$ and, only $W_{12}$ becomes non-zero which provides 
 $E_{\pm}=\pm\la_d\sqrt{c_{2,q}^2c_{1,j}^2+c_{1,q}^2c_{2,j}^2}$.
 Therefore, due to the application of this small perturbation, 
the Majorana modes at each end mix with each other and move away from zero energy 
in pair~\ct{degottardi13}. 
One can also note that if we apply the impurity term deep inside 
the bulk, the edge states do not move away from zero energy, 
since $|c_{ni}|^2$ decreases with increasing $i$.
On the other hand, if the number of Majorana modes at each end is odd, 
one mode always remains at zero energy. In the present situation of off-critical quench, the initial edge Majorana 
interacts with these two non-degenerate states only and oscillates between them that reduces it
to effectively a two-level problem.
The time period ($T_o$) of this collapse and revival is determined by the energy difference 
$\Delta E$ between these two low energy levels ($E_{\pm}$), given by 
\be
T_0= \frac{2\pi}{\Delta E}\propto \frac{1}{\la_d}.
\ee

One can find that the energies associated with these two levels increase linearly with $\la_d$ up to 
a certain value and then there is a saturation in energy (see Fig.~\ref{imp}(b))
for larger values of $\la_d$. This suggests that the time period $T_o$ of collapse and 
revival in the MSP is inversely proportional to $\la_d$ as  observed 
in Fig.~\ref{imp}(a). The above observations lead to a conclusion that 
when an impurity term is suddenly added in the system the initial Majorana
oscillates between two Majorana sites $a_1$ and $a_2$ with a time period 
being inversely proportional to the strength of the impurity. 
This is an interesting observation that even for the off-critical quench the MSP shows
a perfect oscillation consisting of collapse and revival as a function of time.
On the other hand, for higher values of $\la_d$ the MSP exhibits a damped oscillation 
indicating that the initial Majorana decoheres with time (see Fig.~\ref{imp}(a)). 
In these cases, we find that the overlap $\alpha_n$ becomes non-zero even for bulk energy levels. 
As one  increases $\la_d$, the initial edge Majorana couples with 
the more number of interior bulk levels  and hence  the temporal  decay in the MSP is faster 
with increasing $\la_d$.

\begin{figure}[ht]
\begin{center}
\includegraphics[width=3.7in]{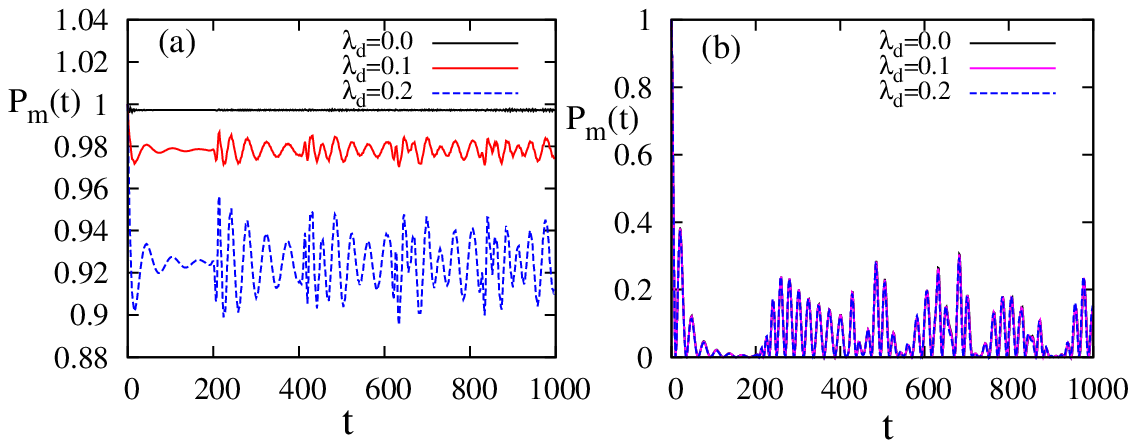}
\vskip -0.5cm
\includegraphics[width=3.7in]{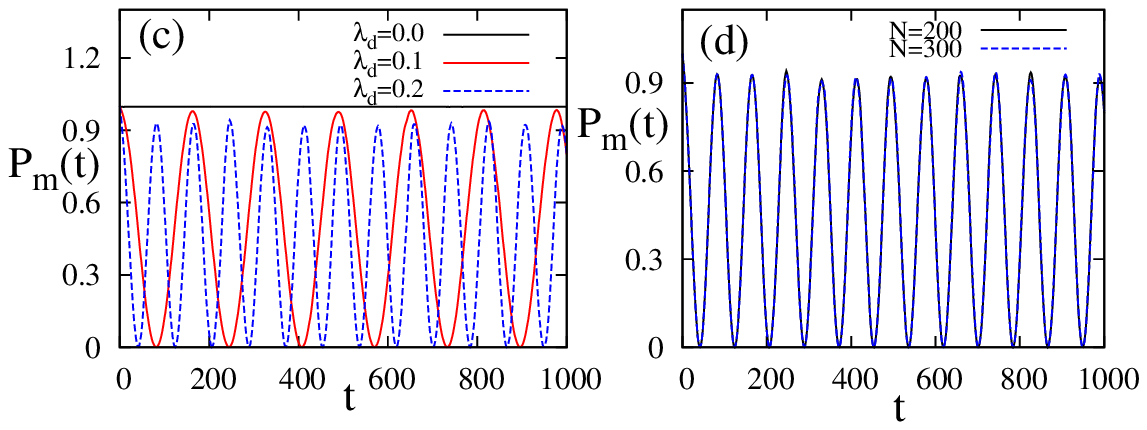}
\end{center}
\caption{(Color online) The MSP as a function of time when the system is quenched from
upper $n=2$ phase ($\la_1=1.8$, $\la_2=3.0$) to $n=1$ phase ($\la_1=2.2$, $\la_2=3.0$) or vice versa. 
 (a) The MSP of $a_1$ Majorana mode decays initially and then becomes rapidly 
oscillating function of time with different mean values depending upon the values of $\la_d$.
(b) On the other hand, the MSP of $a_2$ Majorana mode decays 
rapidly and does not revive significantly for this quenching scheme. 
The plot (c) shows the MSP of $a_1$ mode as a function of time $t$ for smaller value of $\la_d$ when 
the sudden quenching is carried out in the reverse path i.e., 
from $n=1$ phase to $n=2$ phase. We consider $N=200$ for the above three cases. 
(d) This plot indicates that the time period of the collapse and revival of the MSP here
is independent of system size confirming the off-criticality 
of the associated dynamics. Here, $\la_d=0.2$. 
For all the cases we consider $\la_d a_1 a_2$ as the impurity Hamiltonian. }
\label{un2_n1}
\end{figure}

At the same time, we numerically calculate the MSP using the above quenching protocol to study the effect 
of quasi-local impurity, i.e., when $q>2$ in Eq.~(\ref{imp_ham}).
In Fig.~\ref{imp}(c), the MSP shows perfect collapse and revival for different values
of $q$ with $\la_d=0.2$. We find that the logarithm of the time period of these oscillations 
is proportional to $q$ (see Eq.~(\ref{imp_ham}));
$T_o\propto e^{\beta q}$ 
where the factor $\beta$ is different for odd and even $q$.
This form of time period can be explained by analyzing two energy levels (with non vanishing $\alpha_n$) close 
to zero as shown in the inset of Fig.~\ref{imp}(d). This plot shows that the 
energy levels decrease exponentially as a function of $q$ with two different values of
$\beta$ for odd and even $q$; these are in good agreement with that of the obtained from $T_o$.

Let us now focus on the quenching from upper $n=2$ phase to $n=1$ phase and study the MSP 
for different values of impurity strength applied to the final Hamiltonian (see Fig.~\ref{un2_n1}a,b). 
We numerically investigate the dynamics of 
 both $a_1$ and $a_2$ Majorana modes initially 
existed in the upper $n=2$ phase.
In this quenching process the MSP for $a_1$, shown in Fig. (\ref{un2_n1}a), does not decay
rather fluctuates haphazardly with a mean value close to unity. 
This is due to the fact that $a_1$ Majorana mode in the $n=1$ phase  remains 
unaffected in the presence of the impurity term. A pair 
of Majorana modes ($a_1$ and $b_N$) exists at two ends of the chain even on the critical line $\la_2=1+\la_1$ 
separating upper $n=2$ and $n=1$ topological phases which continues to the $n=1$ phase also.
Therefore, in the true sense 
$\la_2=1+\la_1$ is not a phase boundary for the $a_1$ edge Majorana mode. 
It is noteworthy that
the mean value of the MSP decreases as
one increases $\la_d$  in the final Hamiltonian.
On the other hand, the MSP for $a_2$ mode decays rapidly to zero and remains at zero with some irregular 
oscillations for the above quenching process (see Fig. (\ref{un2_n1}b)), since the $n=1$ phase 
does not have $a_2$ Majorana edge mode. The different curves of the MSP for various values of 
$\la_d$ fall on top of each other, thus confirming that the $n=1$ phase remains unaffected 
by the application of $\la_d$.

\begin{figure}[ht]
\begin{center}
\includegraphics[width=3.7in]{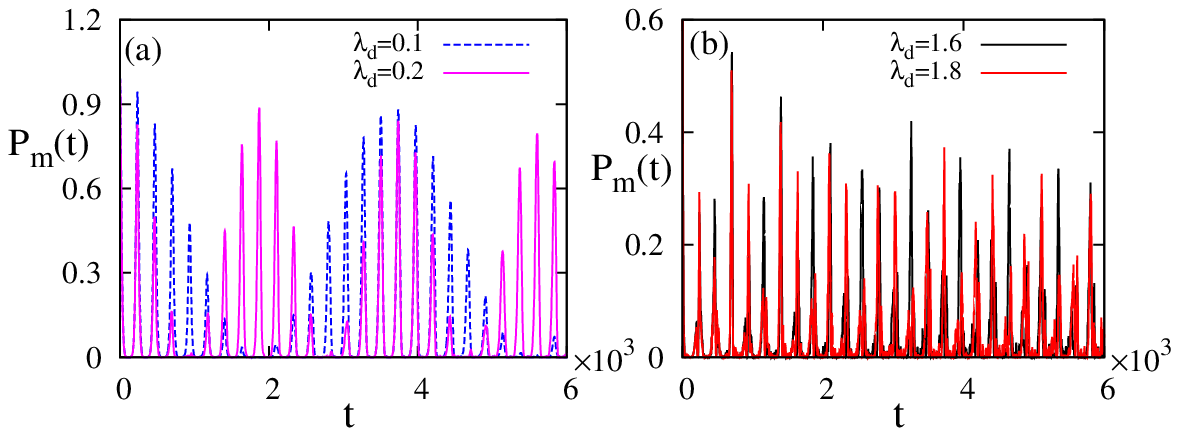}
\vskip -0.5cm
\includegraphics[width=3.7in]{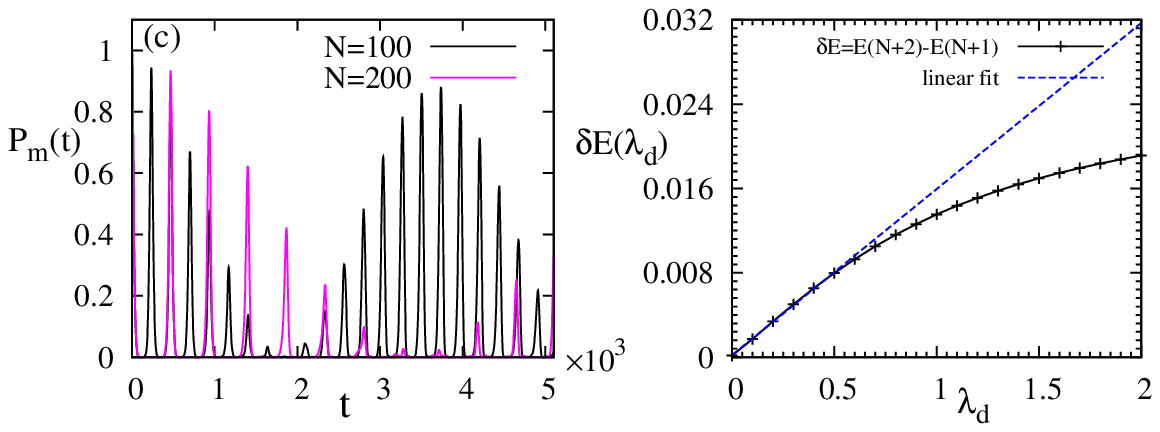}
\end{center}
\caption{(Color online) Variation of the MSP of a zero-energy Majorana mode ($a_1$) 
as a function of time when the Hamiltonian
is quenched from lower $n=2$ phase ($\la_1=1.0$, $\la_2=-1.1$) to the QCP ($\la_1=1.0$, $\la_2=-1.0$) 
lying on $b-d$ line. (a) Beating pattern in the MSP as a function of time when 
the impurity term of small magnitude is applied along with the final Hamiltonian. 
It can be observed that the time-period of the envelope of the beating structure 
varies as $T_{r2}\propto1/\la_d$. (b) On the other hand, for larger values of 
$\la_d$ one can not find any beating pattern in the MSP.
Here, $N=100$ for the above two plots. Figure (c) depicts that the  interior 
frequency of oscillation in the beating pattern is linearly dependent on $N$. Here, $\la_d=0.1$.
(d) Variation of the energy difference between two consecutive energy levels of low 
lying states as a function of $\la_d$.
For all the above cases we consider $\la_d a_1 a_2$ as the impurity term.}
\label{bn2_bdc}
\end{figure}

We now perform a rapid quench following the inverse path as compared to previous one,
i.e., the initial Hamiltonian is in phase $n=1$ while the final Hamiltonian is in 
phase $n=2$ with an impurity term $H_{\rm imp}=\la_da_1 a_2$. 
As shown in Fig.~\ref{un2_n1}(c), the MSP for $a_1$ mode with $\la_d=0$ remains unity  
as a function of time. This is due to the fact that $a_1$ Majorana mode exists in the phase 
$n=2$ when the impurity term is not applied there.
In contrast, we find that the MSP shows collapse and revival with time having
time period $T_o \propto \la_d^{-1}$ when an impurity 
term of smaller strength is applied in the final Hamiltonian. This also can be explained with the same lines of 
arguments as given for Fig.~\ref{imp}(a). Similar to the case of off-critical
quenching inside the same phase, here also,
the MSP displays damped oscillations when $\la_d$ becomes large. The signature of off-criticality of the 
dynamics is depicted in Fig.~(\ref{un2_n1}d) showing that collapse and revival nature of  MSP
is independent of $N$.


\subsection{Critical quenching: MSP }
\label{result2}
In this section, we study the MSP when the system is suddenly quenched up to a 
critical point with an additional impurity term $H_{\rm imp}$ in the final Hamiltonian. 
We shall restrict our focus only on the Majorana mode which is destroyed by 
the application of $\lambda_d$ in phase $n=2$, i.e., $a_1$ and $a_2$ modes that give 
interesting results. 

\begin{figure}[ht]
\begin{center}
\includegraphics[width=3.0in]{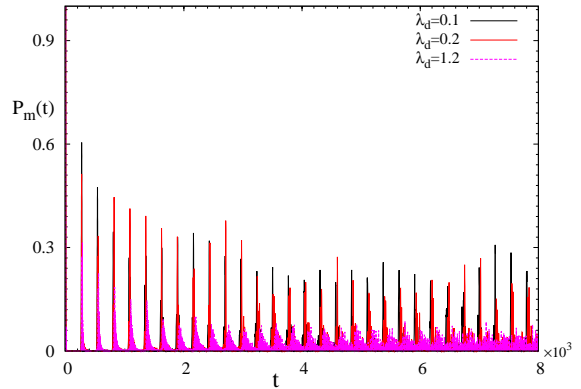}
\end{center}
\caption{(Color online) $P_m(t)$ for $a_1$ Majorana mode as a function of time $t$ for the quenching from 
upper $n=2$ phase ($\la_1=0.2$, $\la_2=2$) to $(n=2)-(n=1)$ phase boundary ($\la_1=1$ and $\la_2=2$) 
for different values of $\la_d$ with impurity term $H_{\rm imp}=\la_d a_1 a_2$ in the final 
Hamiltonian. In this case the MSP does not exhibit any beating pattern and also the revival 
structure destroys with increasing value of $\la_d$.
}
\label{un2_a}
\end{figure}

Let us first investigate the time evolution of the Majorana modes $a_1$ and $a_2$
under the sudden quenching from $n=2$ phase to $(n=2)-(n=0)$ phase boundary designated by $bcd$ line of the phase diagram (see Fig.~\ref{pd}).
As shown in Fig.~\ref{bn2_bdc}(a,c),
MSP exhibits a 
beating like structure as a function of time for the critical quench with 
a small impurity term in the final Hamiltonian. This implies that the system has two 
different energy scales or frequencies in such situation. A close observation of Fig.~\ref{bn2_bdc}(a,c) suggests that
  these two frequencies are related to the system size 
$N$ and impurity strength $\lambda_d$. At the critical point, the energy levels 
close to zero energy are inversely proportional to the system size, i.e., $E\propto N^{-1}$. 
On the other hand, similar to the case of off-critical quench, here also the application of $\lambda_d$ gives rise to another 
energy scale in the system.


We shall now analyze the beating phenomena of the MSP in detail. We have found that there are two types 
of energy differences between two consecutive low lying energy levels  of 
the critical Hamiltonian: one is marked by $\delta E_L$ whose value is larger than 
the other one ($\delta E_S$). We find that the smaller energy difference $\delta E_S \propto \la_d$ 
for a fixed $N$ and $\delta E_S \propto N^{-1}$ for a fixed $\la_d$. On the other hand,
 $\delta E_L$ is very weakly dependent on $\la_d$ rather strongly proportional to $N^{-1}$ 
with changing $N$. For small $\la_d$, these two energy differences remain almost constant 
over a few low-energy levels for which the overlap function $\alpha_n$  becomes non-zero.
As a result, the non-zero contribution in the summation of Eq.~(\ref{prob}) arises due 
to the overlap of the Majorana edge mode with a few low-energy bulk states which have 
two types of equispaced energy levels (with the corresponding consecutive energy differences 
$\delta E_L$ and $\delta E_S$) alternatively. These two constant energy differences is the key 
factor to provide two frequencies in the non-equilibrium dynamics of the system. 
As mentioned already, the relatively shorter time  period of the interior oscillations in the MSP 
is governed by the larger energy difference $\delta E_L$:    
$T_{r1}\sim 2\pi/\delta E_L\propto N$, whereas, the larger time period of 
the envelope is $T_{r2}\sim2\pi/\delta E_S\propto \la_d^{-1}$ or $N$. 
Therefore, we can conclude  that only $N$ controls the interior frequency  and  the
envelope  frequency of the MSP is dependent on both $N$ and $\la_d$.
For the case of $\la_d\to0$, the value of $\delta E_S$ is very small and we find
$\delta E_L/\delta E_S \sim O(10^2)$ leading to $T_{r2}/T_{r1} \sim O(10^2)$.
We find that the time-period of the envelope associated with the beating 
structure for $\la_d\to 0$ is very large compared to the interior oscillation. 

One can intuitively understand the oscillation in the MSP that depends on $N$ for the critical quenching.
Considering periodic boundary condition without the impurity term, the Loschmidt 
echo in momentum representation can be represented as
$ {\cal L}(t)= \prod_{k>0} \left(1 - B_k \sin^2 (E_kt) \right)$, 
where $B_k$ is a very slowly varying positive function of the momentum $k$ ~\ct{happola12,rajak14}. Now as long as 
the dispersion $E_k$ is linearly dependent on $k$ 
near the critical mode, the revival time $t_k= \frac 1 {2} p N |\partial E_k/\partial k|$ for each $k$
becomes independent of $k$, with $p$ being an integer. The MSP is indeed the Loschmidt 
echo with the initial ground state replaced by the initial zero energy edge Majorana state~\ct{rajak14}; therefore, MSP follows the 
similar kind of behavior in the critical quenching. Hence, in addition to the $\la_d$ energy scale which
is also appeared in the off-critical quench, here $N$ dependent energy scale comes into play in the dynamics
resulting in to the beating structure in the MSP.

On the other hand, for larger values of $\la_d$, the overlap $\alpha_n$ remains non-zero 
for more number of low energy bulk modes as compared to the case of small $\la_d$. 
Also, $\delta E_S$ and $\delta E_L$ do not remain 
constant (rather become irregular) for all these bulk energy states for which the overlap $\alpha_n$ is non-zero. 
As a result, we do not get any prominent beating like pattern in time evolution of the MSP 
(see Fig.~\ref{bn2_bdc}(b)).
We have also shown that the energy difference ($\delta E$) between two consecutive low lying energy 
levels varies linearly for small values of $\la_d$, whereas it becomes non-linear for 
higher $\la_d$ (see Fig.~\ref{bn2_bdc}(d)). 
Although, in Fig.~\ref{bn2_bdc}(d), we have shown only one consecutive energy difference, the nature 
of the plot remains almost same for other few low lying energy levels.


 \begin{figure}[ht]
\begin{center}
\includegraphics[width=3.0in]{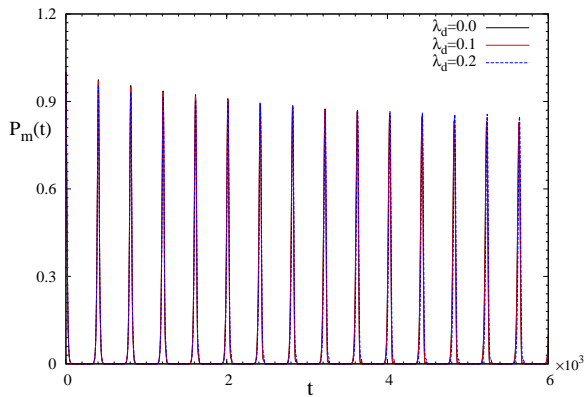}
\end{center}
\caption{(Color online) Time evolution of the MSP of a zero-energy Majorana mode ($a_1$) 
becomes independent of the impurity strength $\la_d$ when the Hamiltonian is quenched from 
$n=1$ phase ($\la_1=1.0$, $\la_2=0.2$) to the QCP ($\la_1=1.0$, $\la_2=0.0$) lying on the $a-d$ phase boundary. 
Here, $N=200$. 
}
\label{n1_n0c}
\end{figure}

Our next aim is to investigate the behavior of MSP under the quenching from $n=2$ phase to $(n=2)-(n=1)$ phase boundary. 
As shown in Fig.~\ref{un2_a}, for this case the MSP of the $a_1$ zero-energy Majorana 
mode  does not show a  beating pattern even for the smaller values of $\la_d$.
There exists mainly one type of consecutive energy difference i.e., $\delta E \sim \delta E_L \sim \delta E_S$. 
We observe that
$\delta E \sim N$ (for fixed $\la_d$)  though it becomes nearly independent on $\la_d$ except for the level nearest to the 
zero energy. 
Hence the interference effect results in a collapse and revival type of behavior rather than a  beating pattern; the time 
period of the revival is linearly proportional 
to $N$.
We also observe that the revival structure of the MSP destroys for larger values of $\la_d$.
With increasing value of $\la_d$, the $\delta E$ changes more rapidly  even for the low-energy levels. 
As a result, the terms of the summation in Eq.~(\ref{prob}) interfere destructively and the revival 
structure of the MSP destroys with increasing $\la_d$.

We also investigate the MSP following the quenching to the $(n=0)-(n=1)$ phase boundary 
($aed$ line of the phase diagram as shown in Fig.~\ref{pd}) starting from $n=1$ phase. Here, we do not observe 
any beating like structure in the MSP as the impurity can not give rise to another new energy scale other than the regular
energy scale determined by $N$.
As shown in Fig.~(\ref{n1_n0c}), we find that the MSP displays a perfect collapse and revival 
with the revival time period $T_r \propto N$. It also has been observed that
the MSP curves for different impurity strengths 
overlap with each other suggesting the fact that the energy levels of the system on the $(n=1)-(n=0)$ phase boundary 
remain unaffected by $\lambda_d$. In this situation, the system
contains only one energy difference $\delta E \sim \delta E_L\sim\delta E_S$ which is inversely proportional to 
$N$ and independent of $\la_d$.

We can now make a comment after a rigorous inspection of all the critical quenching cases 
that the beating like nature is an outcome of the existence of two types of consecutive energy difference 
in the low lying levels namely, $\delta E_S$ and $\delta E_L$. The frequencies of the envelope and interior 
oscillations are dependent on the behavior that how $\delta E_S$ and $\delta E_L$ vary
with $N$ and $\la_d$. On the other hand, the regular collapse and revival of MSP is observed  when there exists 
only one type of energy difference $\delta E$ between consecutive levels; $\delta E$ is inversely proportional 
to $N$ and nearly independent of $\la_d$.

\section{Conclusions}
\label{conclusion}

In summary, we consider a next-nearest-neighbor interacting $p$-wave superconductor derived from 
a three-spin interacting Ising model in presence of a transverse field; we investigate the effect 
of an impurity term in the evolution of the MSP following 
different sudden quenches. It already has been shown that the impurity term that we have considered changes the topological 
nature of the topologically non-trivial phase with two zero-energy Majorana modes \ct{niu12}.
We have found perfect oscillation of the MSP as a function of time when an impurity term of small 
magnitude is applied in the system residing in a topological phase with two Majorana modes at each end.
We attribute this phenomena of perfect oscillation of Majorana, even though the final system 
remains at a off-critical point, to the coupling of the initial Majorana state with only two 
energy levels close to zero energy of the final Hamiltonian. As a result, this eventually 
reduces to the two-level problem.

Additionally, we find that there exists two types of bulk energy difference of two consecutive equispaced low-energy 
states on the $(n=0)-(n-2)$ phase boundary. This leads to a nice beating like 
structure in the MSP when the system is quenched from two edge Majorana phase to the $(n=0)-(n=2)$ phase boundary.
The underlying two frequencies that generate the beating structure are connected with the impurity strength 
and the system size. The beating pattern becomes irregular when the strength of the impurity
is increased appreciably since the Majorana state then starts mixing with higher energy bulk states which are not 
 equispaced.

  
\begin{acknowledgments}
We sincerely thank Amit Dutta and Bikas K. Chakrabarti for fruitful discussion regarding the work.
AR and TN thank IIT Kanpur and SINP Kolkata respectively for providing hospitality during the part of this work. 
\end{acknowledgments}

\end{document}